# Systems Applications of Social Networks

CHANGTAO ZHONG, King's College London
NISHANTH SASTRY, King's College London

The aim of this article is to provide an understanding of social networks as a useful addition to the standard tool-box of techniques used by system designers. To this end, we give examples of how data about social links have been collected and used in different application contexts. We develop a broad taxonomy-based overview of common properties of social networks, review how they might be used in different applications, and point out potential pitfalls where appropriate. We propose a framework, distinguishing between two main types of social network-based user selection – *personalised* user selection which identifies target users who may be relevant for a given source node, using the social network around the source as a context, and *generic* user selection or group delimitation, which filters for a set of users who satisfy a set of application requirements based on their social properties. Using this framework, we survey applications of social networks in three typical kinds of application scenarios: recommender systems, content-sharing systems (e.g., P2P or video streaming), and systems which defend against users who abuse the system (e.g., spam or sybil attacks). In each case, we discuss potential directions for future research that involve using social network properties.

## 1 INTRODUCTION

There has been an explosion of interest in social networks following the popularity of online social networking sites such as Facebook, Twitter and LinkedIn. According to boyd and Ellison [2008], "social network sites are web services that allow individuals to articulate a list of other users with whom they share a connection". The web of personal relationships or "Friendships" is typically viewed abstractly as a social network graph or "social graph" with the site's users as nodes, and each "Friendship" corresponding to an edge between two nodes.

Although social network sites are mostly seen as a way to connect users with other users who are their friends, because these lists are machine accessible, other *systems* can make use of this information as well. For example, recommender systems can use social networks to recommend items to a user based on what her friends like. Such recommendations are based on the assumption that people connected by social links tend to like similar things, a property termed homophily. Similarly, social links are thought to encode a level of trust and legitimacy; thus it has been suggested that the the existence of links and regular communications can be used to separate "trustworthy" identities from suspected spammers.

The goal of this article is to articulate the ways in which novel capabilities such as recommendations and trust can be built into systems using social networks and their properties. To this end, we first list known properties of social networks and categorise them into two types (see Figure 1): *local* and *topological properties*, according to the number of participating nodes. *Local properties* describe features of a single node or relationships of two nodes (i.e., a property of an edge). Such properties can be inferred locally by a node, and therefore, are more amenable to distributed implementations. In contrast, *topological properties* are used to describe structures with more than two nodes, and involve a "neighbourhood" of the social network. Thus topological properties will need to be computed either centrally, or in collaboration with other nodes. Our goal is to aid system designers in choosing appropriate properties that would be useful in their systems. Therefore, we also discuss properties which should be used with caution, discussing limitations and contexts where such properties may not apply.

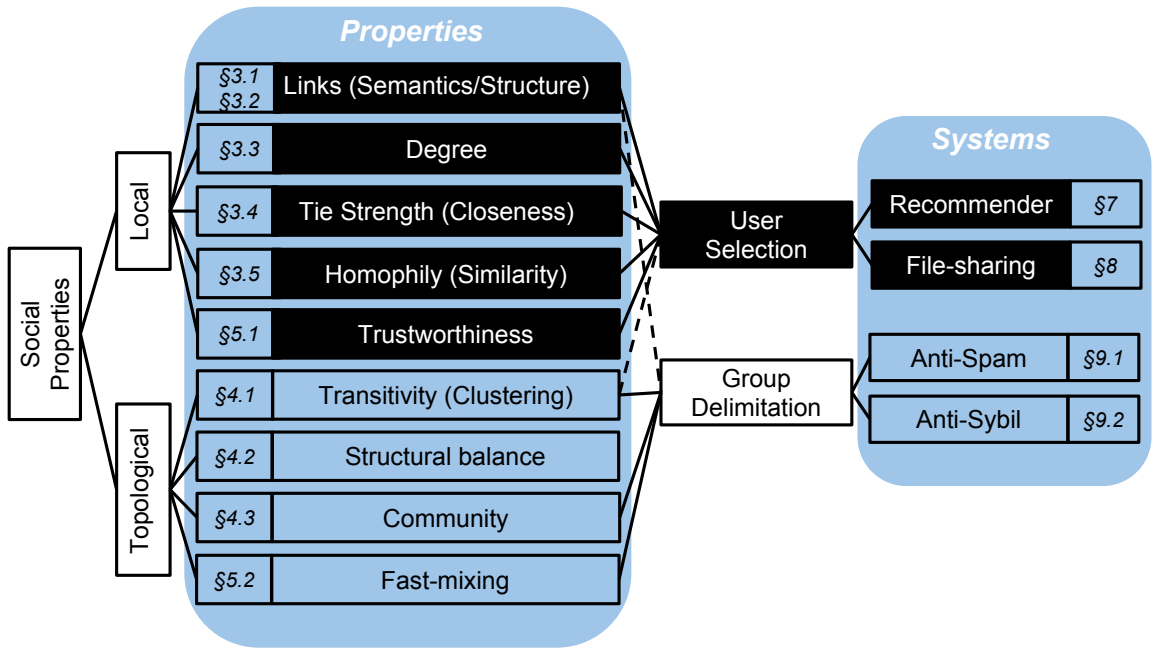

Fig. 1. **Social properties and systems applications.** This figure shows the framework of this survey. In *Properties* part, social properties are categorised into two types: local and topological properties. Then in the *System* part, we survey two methods of utilising social information: user selection and group delimitation, using four kinds of example systems: recommender systems, content- or file-sharing systems and anti-spam or anti-sybil systems.

In the literature, local and topological properties have been used as a means to achieve two ends in application-specific contexts: *user selection* and *group delimitation*. *User selection* involves finding some specific users (termed *target users*) who are "relevant" to another specific user (termed *local users*). The notion of relevance may be based on some application-specific requirements, and the target users are found by exploiting information from social networks. For example, recommender systems may use social properties such as homophily to find a recommending source (other users who can provide clues for generating recommendations) for each local user. Typically, direct relationships between local and target users are considered, and therefore user selection typically involves local social properties. *Group delimitation* involves selecting target users according some general application-specific principles, without considering specific local users. For example, anti-spam systems may wish to identify and blacklist target users who are spammers (or alternatively, identify target users who are not spammers and white list them). The specific requirements to identify target users such as spammers are set by application requirements, and would not change when considering different users. For example, Boykin and Roychowdhury [2005] observed that, by definition, spammers need to send emails to several users, most of whom are strangers to each other and do not normally send emails between themselves. Normally, strong social communities are highly clustered, and resemble cliques, with most people within the community knowing most other people: If Alice communicates with Bob, and Bob communicates with Charlie, Alice and Charlie tend to be friends as well. The level of clustering or cliquishness is measured by a property called "clustering co-efficient". If Mallory is a user with low clustering co-efficient (i.e., he has several correspondents e.g., Dan, Ed, Fred, etc., and most of them are strangers to each other), he may be a spammer.

| Systems | SN type | Main methods | Examples |
|---------|---------|--------------|----------|
| Recommender | Online SN; mainly declared | User selection | To recommend items to a user, SOFIA [Dell'Amico and Capra 2008] selects friends that are similar to her as recommending source. Then items liked by those friends are recommended to the user. |
| File-sharing | Activity SN; mainly inferred | User selection | When a user search content in peer-to-peer systems, Sripanidkulchai et al. [2003] find her social neighbours are more likely to hold the file, so queries are preferentially sent to those neighbours. |
| Anti-spam | Email SN; mainly inferred | Group delimitation | Users with low clustering co-efficient in social networks are consider as suspects, when Boykin and Roychowdhury [2005] try to detect spam in email systems. |
| Anti-Sybil | Online SN; mainly declared | Group delimitation | SybilGuard [Yu et al. 2006] verify suspect users through performing random walks in social networks. |

Table 1. **Social networks and methods utilised in systems applications.**

User selection and group delimitation can be accomplished using both *declared* social networks, where users manually specify social links, and *inferred* social links, where links are automatically deduced based on user actions. To provide a flavour for how different social properties can be used, we study four kinds of social applications (Table 1): recommender systems (user selection, typically using declared links), file-sharing systems (user selection typically over inferred links), anti-spam systems (group delimitation using inferred links) and anti-sybil systems (group delimitation typically using declared links). The survey itself is divided into two parts. In the first part, we survey different kinds of social networks which are available to systems designers, and the properties which most social networks can be expected to possess. In the second part, we survey social applications as outlined above.

## PART I: SOCIAL NETWORKS AND THEIR PROPERTIES

In the first part of this survey, we review social networks and their properties. In §2, we review examples of social networks, with a view to illustrate how social network data has been gathered or inferred in the past. Then, in §3 and §4, we discuss how social properties have been used in systems applications. Finally, we propose a framework for understanding the application of social properties in a systems context (§6, Figure 1).

## 2 EXAMPLES OF SOCIAL NETWORKS

In recent years, there has been an explosion of interest in social networking sites such as Facebook, Twitter and LinkedIn. Whilst there have been older avenues for online social interaction, such as Ward Christensen's Computerised Bulletin Board System or CBBS (1978) and Richard Bartle and Roy Trubshaw's original Multi-User Dungeon or MUD (1977)[1], the current social networking phenomenon is distinguished by the massive scale of its adoption in the general population. For example, Facebook claims to have 1.28 billion daily active users on average for March 2017 [Facebook 2017]. A large amount of activity happens on social networks. In 2008, Leskovec and Horvitz [2008] analysed 30 billion instant messaging conversations among 180 million users over the course of a month. Thus, modern social networking has generated extremely large social graphs.

In this section, we summarise early attempts at analysing different networks, and where applicable, discuss how the social relationships were reconstructed or gathered, as well as how this data was utilised in systems design. Rather than an attempt at cataloguing all available social networks (of which there are too many to mention), this section aims to provide a window into the diverse varieties of inferred and user-declared user–user connections which have been examined or utilised, and may inspire systems designers for their own applications.

## 2.1 Pre Web 2.0 online communities

Before Web 2.0, online communities typically did not store an explicit social graph. Thus many graphs in older online communities are *inferred* relationships based on records of interactions. Studying such systems is especially valuable to designers of legacy systems who may wish to incorporate social features in later versions. Cobot was a software agent written to gather social statistics from LambdaMOO, a text-based online virtual world[Isbell et al. 2000]. Sharing social statistics with other (human) participants in LambdaMOO improved social interactions, and engendered new communication topics. Even earlier, Schwartz and Wood [1993] looked at the social graph induced by email conversations and deduced shared interests. Leskovec and Horvitz [2008] examined the network of instant messaging conversations and found strong evidence for homophily.

ReferralWeb [Kautz et al. 1997] synthesized a social network by mining a variety of public sources including home pages, co-authorships and citations, net news archives and organisation charts, and used this to effectively locate experts for different topics. This technique may be useful in other systems as well: In many applications, including at the systems level, it may be relevant or useful to capture multiple modes of relationships or interaction. For example, email and instant messaging, when combined with phone call logs, could yield a more complete picture of interaction between two individuals, than can be seen by examining each mode of interaction separately. Some researchers also constructed multilayer social networks to represent such information, which we will discuss later in §2.4.

One early example of *explicitly* stored social graphs comes from webs of trust. Prominent early webs of trust include PGP [Stallings 1995] and Advogato [Levien 2000]. However, these are usually stored in a decentralised fashion, and may be difficult to capture completely. More centralised records of trust and reputation relationships may be found on online shopping sites, where trust webs have been exploited to enhance confidence in sellers and items, as well as to provide useful recommendations. Many such networks have been examined for inferring relationships between users and their consumption patterns. Examples include Epinions [Richardson and Domingos 2002], Amazon [Leskovec et al. 2007] and Overstock [Swamynathan et al. 2008].

## 2.2 Modern social networks

One of the earliest studies looking at recent Web 2.0 online social networks was conducted by Mislove et al. [2007]. It showed that online social networks are scale-free, small world networks and have a power-law degree distribution. It also identified a form of reciprocity: in-degree distributions match out-degrees. Fu et al. [2008] studied a Chinese social network (Xiaonei, now known as Renren[2]) and a blog network and showed that social networks are assortative whereas blogs are dissortative. Other prominent Web 2.0 communities analysed include the Flickr photo sharing community [Cha et al. 2009b] and video sharing on YouTube [Cha et al. 2009a].

---

[1]For a brief and vividly sketched history of Social Networking, see The Cartoon History of Social Networking [Lon11 2011]. Other brief accounts with similar material include [Bianchi 2011; Nickson 2009; Simon 2009].

Some networks such as Facebook and Twitter have received considerable attention because of their popularity or importance. These studies have provided evidence for various social process. For example, Sun et al. [2009] show that information contagion, in terms of people "liking" pages on Facebook, starts independently as local contagions, and becomes a global contagion when different local contagions meet. Similarly, Wilson et al. [2009] and Viswanath et al. [2009] showed that the network of nodes with interactions between them is much more sparse than the graph of declared relationships. We discuss this further in §5.1.1, in relation to the trustworthiness of interaction-based links. Cha et al. [2010] discuss three different measures of user influence on Twitter. Kwak et al. [2010] performed a large-scale study of Twitter and found low levels of reciprocity, unlike other social networks. Instead they found that after the first retweet, most tweets reach around 1000 people, regardless of where it started; and that URLs are mostly news URLs. Based on these findings, they proposed an alternate view of Twitter as a news media site rather than a network for just social interactions.

## 2.3  Synthetic social networks

The data of many social networks is not easily obtained. Alternately, the terms of service of a social network may not allow the offline use and storage of their users' data. Thus, during development of social network-based systems, it may be desirable to use synthetic but realistic social graphs. Synthetic graph generators have been designed for Facebook [Sala et al. 2010] and Twitter [Erramilli et al. 2011]. Older methods for synthetic social graphs were built upon generative models which capture particular properties of social networks. For example, Barabasi's Preferential Attachment model captures scale-freeness and power-law degree distributions (§3.1). Graphs generated according to the Kleinberg or Watts-Strogatz model capture the small world phenomenon (§4.1). See [Chakrabarti and Faloutsos 2006] for a survey of different generators. More recent generators include ones based on Kronecker multiplication, which capture densification laws and shrinking diameters [Leskovec and Faloutsos 2007].

## 2.4  Multilayer social networks

Sociologists have recognised that there is more than one method to constructing social networks [Krackhardt 1987; Padgett and Ansell 1993; Wasserman and Faust 1994] and people construct their profile identities differently on different platforms [Zhong et al. 2017]. So [Kivelä et al. 2014] argue it is often "an extremely crude approximation of reality" to represent social systems using a network with only a single type of relationship. Thus, it is crucial to construct multilayer social networks using different types of ties among the same set of individuals to study social systems [Scott 2012; Wasserman and Faust 1994]. For example, Szell et al. [2010] identify six different types of one-to-one interactions between players: friendship, communication, trade, enmity, armed aggression and punishment. They construct a multilayer social network for each kind of interaction and find that the social structures in each layers are quite different. Zhong et al. [2014] examine the overlapping social links between Pinterest / Last.fm and Facebook. They find that although a large fraction social links are copied from established networks to new websites, active and influential users tend to wean from copied social links to interact more over links created natively. See [Kivelä et al. 2014] for a survey of different multilayer social networks.

---

## 3 LOCAL PROPERTIES

Although the complete social graph may never be articulated online in its entirety, it is thought to possess certain characteristic features, some of which have even become catch-phrases in popular culture. Many of these properties are not unique to social networks; rather they are common to other "complex networks" [Albert and Barabasi 2002; Boccaletti et al. 2006; Newman 2003b; Vespignani 2005]. It should be noted that graph properties can evolve over time. For instance, it has been observed that the number of edges grows superlinearly in the number of nodes, creating denser graphs, and the diameter of the graph often shrinks over time [Leskovec et al. 2007]. In this and following two sections, we summarise some well-known properties of social networks, and discuss how they can be utilised in systems applications.

We divide "social properties" into two classes:

**Local properties**  We consider those properties that describe features of a single node or a couple of nodes as local properties. For example, node degree is a node-centric property, which shows the number of other nodes the local node connected with. It describes features of a single node and will be different for different nodes. Some other properties are edge-centric properties, which describe properties of a pair of node. These properties include link semantics, link structure and tie strengths.

**Topological properties**  We consider those properties that describe features of more than two nodes as topological properties. For example, the transitivity shows whether friends of friends can be friends, while structural balance describe the stableness of structures within signed social networks. Some other properties also describe global features of social networks, such as community structure, which divides the whole network into different densely connected sub-networks.

We introduce local properties of social networks in this section, and topological properties will be described in §4.

### 3.1 Properties of node degrees

Social graphs are thought to be *sparse* [Girvan and Newman 2002]: most node pairs do not have direct edges between them. The degree of nodes, the number of other nodes to which a node is connected, typically has a *right-skewed distribution* [Girvan and Newman 2002]: the majority of nodes have a low degree but a few nodes have a disproportionately large degree. The precise distribution often follows a power-law or exponential form. Thus in social network analysis, it is quite common to use the degree of nodes (or *degree centrality*) to show which are the most important nodes [3].

But social graphs differ from other networks in having positive correlations between node degrees [Newman 2002]. In other words, nodes with high degree tend to connect to other nodes with high degrees, and nodes with low degree tend to interconnect among themselves. This property is known as *assortativity*. By contrast, biological and technological complex networks tend to be dissortative, i.e., there is a negative correlation of node degrees, with high degree nodes preferentially connecting to low degree ones. Systems designers should note that assortative networks percolate more easily than dissortative networks and are also more resilient to targeted attacks which remove high-degree nodes [Newman 2003a].

---

[3]Some other metrics can also be used in such centrality measurement. One another example is *betweenness centrality*, which equals to "the number of shortest paths pass through a node from all nodes to the node"[Freeman 1977]. A node with high betweenness centrality has a large influence on the transfer of items through the network.

## 3.2 Link semantics

At a semantic level, differences arise between different social networks because they capture different kinds of relationships between their users. Facebook, for example, tries to recreate the friendships of individuals, whereas LinkedIn attempts to capture their professional contacts. Still others, such as the educational social network, Rafiki [4], encourage and foster relationships between teachers and students across the world. Geospatial social networks like Foursquare engender connections between people with ties to similar places. Interest-based social networks like the epinions online community, Pinterest, last.fm social network etc. tend to form links between people who like the same items or the same kind of items.

It is also possible that links in a single social network have different semantics. For example, in Pinterest social network, one user may befriend another user due to matched interests in one type of images, say images related to "food and drinks". At the same time, the user may have other interests, such as "fashion" or "pets", because of which they may befriend another user.

**Implication**: Clearly different social networks are suitable for different needs. For example, in [Sastry 2012], we use interest-based social networks to capture affinities between people and content that interests them. In [Sastry et al. 2011] we use an entirely different network, based on human contact patterns, and find opportunities to deliver data over a series of such contacts.

Link semantics can also used to distinguish different kinds of links inside a single social network. One example is to use it for distinguishing different kinds of friends of a user. For example, Yang et al. [2012b] develop a recommender system based on the idea, and take the semantics of different social links into consideration when recommending different items.

## 3.3 Link structure

At a structural level, the friendship edges can be drawn in a variety of ways. Most social networking sites require users to explicitly declare their friends. Some, such as Facebook and LinkedIn, have bidirectional (or equivalently, undirected) edges which are formed only when both parties confirm their friendship to the site. Others, such as Twitter, have unidirectional (directed) edges in which one user can "follow" another user without the other user's permission. Unidirectional links are particularly suited for interest-based social networks where reciprocation is not expected or required. For instance, Digg and vimeo allow users to add other users as contacts and unilaterally subscribe to their content uploads and recommendations.

Social graphs can also be inferred automatically (implicitly) as a side-effect of user actions. Human contact networks draw edges between people who are in close proximity to each other [Eagle and Pentland 2005; UCSD 2004]. Other examples include email networks, where edges are drawn as a result of email interaction. Note that automatically inferred links still meet boyd's generic definition of social networks [boyd and Ellison 2008] because the links are still explicitly articulated in a machine accessible manner.

**Implication**: Bidirectional links are more trustworthy because they are more difficult to make: spammers or other nodes can indiscriminately make unidirectional links to several unrelated nodes whereas bidirectional links have to be vetted by both parties. However, the directionality of unidirectional links can be used to create a reputation mechanism similar to PageRank [Page et al. 1999]. TunkRank [Tunkelang 2009], developed as a direct analogue of PageRank, measures user influence in Twitter based solely on link structure. TwitterRank [Weng et al. 2010] gives a topic-sensitive ranking for Twitter users, using a random surfer model similar to PageRank. MailRank [Chirita et al. 2005] applies methods similar to PageRank to rank email addresses, so that regular users and spammers could be separated from each other. IARank [Cappelletti and Sastry 2012] is a model that identifies influential Twitter users according to their capacity to increase the spread of tweets through retweets or mentions.

---

[4]Website has been down according to http://rafikionline.wordpress.com/about/ (last accessed Jun 15, 2017)

Both explicitly declared and implicitly inferred social graphs can suffer from irregularities such as missing or spurious links. In explicitly declared social networks, these errors arise due to user actions. When links are automatically (implicitly) inferred, the irregularities could arise due to technology limitations. For example, one way to infer a human contact network is to use Bluetooth device discovery to identify users who are close to each other, and drawing a link between them. This technique was used to collect the MIT/Reality data set[Eagle and Pentland 2005]. However, to conserve batteries, device discovery was only initiated once every five minutes. Therefore contacts shorter than this interval could be missed. Similarly, the ten metre range of Bluetooth along with the fact that it can penetrate some types of walls, means that spurious links can be created between people who are not physically proximate or co-located in a room.

### 3.4 Tie strength

*Tie strength* describes the other aspect of relationships of actors connected by social links. It is found that loose acquaintances, known as *weak ties*, can help a person generate creative ideas [Burt 2004], search useful knowledge [Hansen 1999] or find a job [Granovetter 1995] and help platforms bond users [Zhong et al. 2016]; trusted friends and family, called *strong ties*, can affect emotional health [Schaefer et al. 1981] and often join together to lead organisations through times of crisis [Krackhardt and Stern 1988]. These two types of ties are different in several aspects. For example, friends with strong ties communicate more often and through various different media [Haythornthwaite 2001]. This will end up to different kinds relations between them. But weak ties tend to be the means for information to cross from one community to another, and are thus critical for information diffusion [Brown and Reingen 1987]. In practice, the degree of overlap of two individuals' friendship networks (usually termed *closeness*) is usually used to show the strength of ties [Gilbert and Karahalios 2009; Granovetter 1973; Kossinets and Watts 2006].

**Implication**: Compared with online social networks, measuring tie strength in real life is much more difficult. Recently, technologies like Infrared [Choudhury and Pentland 2003; Groh et al. 2010], RFID (Radio-frequency identification) [Brown et al. 2014; Stehlé et al. 2011] and Bluetooth have been applied to estimate on-going social interactions through mobile devices. For example, Banerjee et al. [2010] train regression models upon RSSI (Received Signal Strength Indicator) values from Bluetooth and WiFi to estimate relative distance and spatial locations of two people. Similarly, Matic et al. [2012] detect social interactions by combining proximity detection results from mobile communication systems and movement speed information from an external accelerometer attached to user's chest. Palaghias et al. [2015] present one of the most sophisticated attempts to capture social interactions. With the help of a hierarchical machine learning method, they estimate distances and relative orientation among users. Then, they allow users to collectively exchange sensed information and divide people they encountered into 3 groups: public, social and personal. In this way, a social network with tie strength information is built [Palaghias et al. 2016].

### 3.5 Homophily, segregation and influence

Homophily, the tendency of people to form links with other people having similar attributes as them, is a powerful driver for social link formation [McPherson et al. 2001]. Studies have shown that rising similarity between two individuals is an indicator of future interaction [Crandall et al. 2008]. Systems can use this finding to optimise for links that are about to form.

Homophily can also lead to segregation. The Schelling model shows that global patterns of segregation can occur from homophily at the local level, even if no individual actively seeks segregation [Schelling 1972]. Thus, homophily could be used to effectively partition data.

[Aral et al. 2009] suggest that peer influence in product adoption (i.e., viral marketing) may be greatly overestimated, and that homophily can explain more than 50% of perceived behavioural contagion. Hence, homophily could be used to explain and anticipate product adoption or data consumption patterns.

## 4 TOPOLOGICAL PROPERTIES

In these section, we introduce some topological properties which describe the relationship among more than two nodes.

### 4.1 Transitivity of friendships and small worlds

One finds a larger than expected (when compared with random graphs) number of triangles, or triads, in the social graph[5]. In sociology, this property is often termed *network transitivity*, because it implies that friendship is transitive: friends of friends are also likely to be friends.

Triangles are the smallest possible cliques (3-cliques) in a graph. Therefore, the excess of triangles can also be seen as evidence of "cliquishness" in a local neighbourhood. Each edge in a triangle is a *short-range connection* between two nodes which are already close to each other (because of the two hop path through the third node in the triangle). Hence, in complex networks literature, this property is usually referred to as *clustering*, in analogy with regular lattices, where edges are formed between nodes which are close to each other.

Despite the lattice-like clustering and the large number of short-range connections, several complex networks, including social networks have been shown to have a small characteristic path length and small diameter, like random graphs. Popular culture knows this as the concept of "six degrees of separation", after Stanley Milgram's famous experiment [Travers and Milgram 1969], and the phenomenon of short paths has been called the *small world effect*.

### 4.2 Structural balance

While most social networks only capture positive relationships or friendships, it is also possible to add negative or antagonistic relationship edges. For instance, users in the Slashdot[6] community can mark each other as friend or foe [Kunegis et al. 2009]. Users in Essembly, an online political forum, can label each other as friends, allies or nemeses [Hogg et al. 2008]. Guha et al. [2004] were the first to examine *signed relationships* in online social networks, when they considered trust and distrust in the Epinions community.

Unlike positive relationships, the presence of antagonistic or negative relationships can induce stresses in the social graph, which make certain configurations of relationships unstable. The theory of *structural balance* governs the different relationships possible. There are two variants. The structural balance property originally proposed by Heider [Heider 1946] holds if in every triad, either all edges are positive, or exactly one edge is positive. Davis [Davis 1967] proposed a theory of weak structural balance which allowed all combinations except for a triad with two positive and one negative relation. Figure 2 shows the possible triads and the social psychology-based motivation

---

[5]Here we restrict our focus to triads. However, it has been extremely useful to generalise this to other local patterns which may occur in graphs, including signed versions. Such patterns, termed as network motifs [Milo et al. 2002], have been useful in classifying different networks. For instance, a study of the relative abundance of different kinds of triads shows that social networks and the WWW may belong to the same "superfamily" of networks [Milo et al. 2004].
[6]http://www.slashdot.org, last accessed on Jun 15, 2017

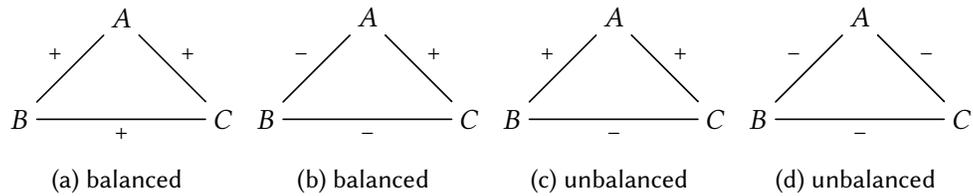

<div align="center">(a) balanced      (b) balanced      (c) unbalanced      (d) unbalanced</div>

Fig. 2. **Possible triad configurations in a signed network, and their balance statuses.** (a) This represents three nodes who are mutual friends. This is stable, and can be seen as a case of "a friend of my friend is also my friend." (b) This represents the case when two users have a mutual enemy, and is also stable. This is a case of "enemy of my enemy is my friend.". The other two labelings introduce psychological "stress" or "instability". In (c), $A$'s friends $B$ and $C$ are enemies. This would cause $A$ to persuade $B$ and $C$ to become friends, reducing to case (a), or else, to side with one of them, reducing to case (b). (d) has three nodes who are mutual enemies. This is also unstable, because there is a tendency for one pair to gang up against the other. However, this tendency is lesser than the stress induced in case (c); hence weak structural balance considers this to be balanced.

for their classification as stable or unstable. Refer [Tang et al. 2016b] for survey of studies on signed social networks.

### 4.3 Community structures

*Communities* are subsets of nodes which are more "densely" connected to each other than to other nodes in the network. Community structures are a fundamental ingredient of social networks. Newman and Park [2003] showed that both transitivity of friendships (clustering) and assortativity can be explained by a simple model that considers social networks as arising from the projection of a bipartite graph of nodes and their affiliations to one or more communities.

Unfortunately, there is no clear consensus on how to define a community, and various measures have been proposed [Flake et al. 2002; Girvan and Newman 2002; Newman and Girvan 2004; Radicchi et al. 2004; Wasserman and Faust 1995]. Nevertheless, community detection is highly central, and the topic of several surveys [Danon et al. 2005; Fortunato 2010; Newman 2004]. While the quality of the different algorithms, in terms of meaningful communities detected, can vary from case to case, most of them are computationally intensive and do not scale to large networks. The Louvain method [Blondel et al. 2008] is a fast algorithm based on Newman and Girvan's modularity measure [Newman and Girvan 2004] that has good scalability properties.

Practitioners should be aware that the Louvain method (and other modularity-based algorithms) can produce inconsistent results depending on node orderings [Kwak et al. 2009]. Leskovec et al. [2010] compare different objective functions for defining communities and the quality of communities found by some common algorithms. Using graph conductance as a measure, Leskovec et al. [2008] show that at small scales ($\approx$ 100 nodes), tight communities can be found (i.e., the best communities of small sizes are separated from the larger graph by a small conductance cut), but at larger scales, the communities "blend" into the larger network (i.e., the community's cut has large conductance).

## 5 "PROPERTIES" WHICH SHOULD BE USED WITH CAUTION

In this section, we discuss two "properties" of social networks, which are widely believed to be true, but recent evidence is emerging which suggests otherwise. Both properties are related to security, and their not holding will negatively impact systems built on social networks.

## 5.1 Trustworthiness of links

Most social systems implicitly make the assumption that users only form links with other users they trust, and further, that correct meanings are attached to the links. Even if users can be trusted to be honest, relying on the correctness of links requires an assumption about the hardness of *social engineering* attacks, which aim to fool honest users into creating false relationships. The security firm Sophos demonstrated the surprising effectiveness of such attacks. It created two fictitious Facebook users, complete with a false profile and an age-appropriate picture[7]. Facebook friend requests were sent out to 200 Facebook users, and nearly 50% of them voluntarily created a link [sophos 2009]. Similarly, Burger King ran an advertising campaign called Whopper Sacrifice, in which it was able to induce a number of users to *shed* 10 Facebook friends each, in return for a free burger [Wortham 2009].

A powerful class of social engineering attacks makes use of the tendency of people to trust their friends or friends of friends. Bilge et al. [2009] describe attacks based on cloning data from a victim's online profile and sending friend requests (either on the same social network where the cloned data was obtained, or on a different social network, where the victim is not yet registered) to the victim's friends. Boshmaf et al. [2011] show that it is easy to expand from one attack edge: Friends of an initial victim are more likely to accept friend requests from an attacker (i.e., attacks are more successful with new victims whose friends are "friends" of the attacker). Disturbingly, these attacks have been shown to have a success rate > 50%.

A similar problem arises with the correctness of node identities. To be sure, many social network operators would like users to have identities that can be relied upon: Facebook's Zuckerberg famously wants people to have a single identity online and offline [Kirkpatrick 2010, p. 199]. Rafiki presents an extreme example of ensuring the integrity of identities. Because it explicitly targets school children, Rafiki has moderators who ensure the integrity of identities, and even go to the extent of monitoring interactions to make sure that the guidelines of the site are not violated. However, such support is not available or feasible on most social networking sites. False identities by themselves are not usually of much concern as long as they are not able to make real users link to them. But attackers do not need fake identities if they can compromise real users. A compromised user automatically gives the attacker access to her "real" links into the social network. The reality of this threat can be seen from numerous reports of compromised profiles of honest users on different social networks [McMillan 2006; Zetter 2009].

*5.1.1 Activity networks: a potential solution considered.* It has been argued that all links on a social network do not represent the same level of trust. For instance, some users might find it difficult or socially awkward to refuse friendship requests and form links out of "politeness". One point of view [Wilson et al. 2009] holds that different social links represent different levels of trust, and systems requiring a high degree of trust should not use all explicitly declared links. Instead edges should be filtered based on the presence of social activity (e.g. bidirectional conversation records).

Unfortunately, while the studies in [Chun et al. 2008] show that such *activity networks* have qualitatively similar topological characteristics as the original social networks, the evaluations in [Wilson et al. 2009] show that systems such as RE: [Garriss et al. 2006] and SybilGuard [Yu et al. 2006] which depend on links being trustworthy, do not perform as well when using the sparser and fewer edges of the activity networks, as they do on the denser native social graph. A further complication is that the activity graph is highly dynamic, with the strength of interaction between node pairs changing over time [Viswanath et al. 2009].

---

[7]21 year old "Daisy Felettin" and 56 year old "Dinette Stonily", whose names are anagrams of "false identity" and "stolen identity" respectively. This expands on a 2007 study by the same firm, using a profile name "Freddi Staur", an anagram of "fraudster".

A different point of view suggests that people may be interacting more often than suggested by records of actual conversations. Benevenuto et al. [2009] study clickstream data of user actions on online social networking sites and find that 92% of all activity on social networks involves browsing friends' profiles. A similar conclusion is reached in [Jiang et al. 2010], which mines the extraordinarily rich and accessible records of the Renren social network (These records include visitor logs to each profile, etc.).

Thus one potential solution for verifying trustworthiness of links could be to use networks inferred from consistent and bidirectional interactions, either based on silent browsing, or explicit conversational activity. However, even this may not work: The social success of bots such as Cabot on LambdaMOO [Isbell et al. 2000] suggests that spammers might be able to fool users even when interaction is required.

**Implication**: From the viewpoint of a systems designer, the weakness of online identities and links implies that systems have to be designed defensively to be robust against compromised or false identities and links, even though most social network links encode a high-level of trust between the human actors involved (especially when they are bidirectional). Some systems, like Sybilguard [Yu et al. 2006] and SybilLimit [Yu et al. 2010] parameterise the security they offer based on the trustworthiness of links. For instance, SybilLimit guarantees that no more than $O(\log n)$ false (Sybil) nodes will be accepted as long as the number of *attack edges*, between a compromised node and an honest node, is bounded by $\Omega(\sqrt{n} \log n)$. It may be possible to leverage trust computations from one domain to reason about trustworthiness in another domain, thereby allowing newer social networks and applications to bootstrap trust from more mature social networks and longer user histories to determine user trustworthiness [Venkatadri et al. 2016].

## 5.2 Fast-mixing

As we will discuss in §9.2.1, many social systems (e.g. [Danezis and Mittal 2009; Lesniewski-Laas and Kaashoek 2010; Yu et al. 2010, 2006]) have been built on the assumption that social networks are *fast-mixing*. This is a technical property which essentially means that a random walk on the network will converge to its stationary distribution after very few steps, i.e., every edge in the graph has an equal probability of being the last hop of the walk, or equivalently, the probability of the walk ending on a node is proportional to its degree[8]. If the mixing time of a graph is $T$, then after $T$ steps, the node on the random walk becomes roughly independent of its starting node. Typically, an $n$-node graph is considered to be fast-mixing if $T = \Theta(\log n)$.

The first use we have come across of fast-mixing in social systems is in the design of Sybilguard [Yu et al. 2006]. There the authors assert that social networks are fast-mixing, citing [Boyd et al. 2006][9], who in turn note that stylised social network models such as [Kleinberg 2000], which are regular graphs with bounded degree, are fast-mixing. Another theoretical proof comes from [Tahbaz-Salehi and Jadbabaie 2007] who show using spectral arguments that the mixing time of a *cycle* graph decreases from $\Omega(n^2)$ to $O(n \log^3 n)$ when shortcut edges are added with a probability $p = \epsilon/n$ (i.e., every one of the $n(n-1)/2$ edges is chosen as a shortcut with a probability $p$). These shortcuts effectively convert the cycle into a small world graph similar to the Watts-Strogatz model discussed earlier. However, this result is higher than the $O(\log n)$ mixing time assumed by the systems mentioned above.

---

[8]This terminology derives from the theory of Markov chains, where the term *mixing time* is used to denote the speed with which a Markov chain converges to its stationary distribution.

[9]Actually, the conference version [Boyd et al. 2005] is cited, but this does not directly talk about social networks.

Unfortunately, the theoretical results about fast mixing do not always generalise to real-world scenarios: Nagaraja [2007] simulate the Kleinberg model and a social network drawn from Live-Journal, and showed that they mostly have good mixing properties, worse than but close to that of expander graphs. However certain parameter choices, such as a large density of short-range links in the Kleinberg model, are shown to severely affect convergence times. Mohaisen et al. [2010] conduct an extensive survey of mixing times in different social network data sets, and finds two sources of variation in convergence times: First, within a given social network, random walks starting from some nodes converge more slowly than others. Since mixing time, by definition, is the maximum of the different convergence rates, this implies slow mixing. Second, some social networks mix more slowly than others. The authors observe that networks such as scientific co-authorship networks, which may encode high degrees of trust, are slower. They note that mixing times improve dramatically if the lowest degree nodes are deleted from the social network, as done in the evaluation of Sybilguard and similar systems.

**Implication**: The empirical results above seem to suggest that fast-mixing may not be a universal property that applies in all social graphs. For *some* nodes, such as those of large degree, or in some densely connected social networks, it may be possible to use random walk-based methods that rely on converging fast to a stationary distribution, but convergence times should be explicitly tested before applying such methods.

Note that the correctness of random walk-based techniques is guaranteed as long as the walk is long enough to ensure convergence to the stationary distribution. Thus, as long as one is willing to pay the performance penalty of long random walks, such systems can still be used. In other words, a weaker property that can always be relied upon is that a sufficiently long random walk in social networks will always lead to a node which is independent of the initial node. This property is guaranteed to hold as long as the social graph is connected and non bipartite [Flaxman 2007].

Unfortunately, the security guarantees of random walk-based Sybil defenses like Sybilguard may be compromised if the walk ends up being too long, and thereby allows walks starting from honest nodes to "escape" to a region controlled by Sybils.

## 6  A FRAMEWORK FOR USING SOCIAL PROPERTIES IN SYSTEMS APPLICATIONS

While the above properties may be used creatively in many ways within specific application contexts, we observe two main modes of usage: user selection and group delimitation. Furthermore, as shown in Figure 1, the social properties used in each of these two usage modes tends to be largely disjoint, i.e., social properties (with a few exceptions) can be split into two depending on whether they are mainly used for user selection or group delimitation. Below, we explain this framework in more detail, and how social properties are used within this framework.

### 6.1  User selection

User selection is a *personalised* method of using social properties in systems applications. The aim is to find some specific users (termed *target users*) who are relevant (in a personalised way) to another specific user (termed the *local user*). The target users are identified based on application-specific requirements, and information from the social network is used to select the target users. Thus the relationship between the local and target users, and their relative (mainly local) properties matter for user selection; and user selection usually identifies target users using local properties (see Figure 1):

**Links (semantics/structure)**  A common requirement is to filter potential target users, who are typically linked with a given local user through uni- or bi-directional links, based on the kind of the link between the local and target user. For example, Victor et al. [2009a] consider a user's immediate friends in the social graph as the recommending source, and recommends items liked by friends to users.

**Degree (centrality)** The degree of nodes can also be used to select target users, for example based on a cutoff threshold. Degree also indicates their centrality, and degree or other forms of centrality can be used to prioritise or select certain nodes. For example, in delay tolerant networks, to increase the probability of reaching the message destinations, a common strategy is to select users with high degree (or other centrality measure) as target users, and forward messages [Hui et al. 2008].

**Tie Strength (closeness)** In general, links may be classified as strong or weak, and users connected by links with a specific strength (according to system requirements) can be selected as target users. In practice, strength is usually evaluated by closeness (number of overlapping friends) [Granovetter 1973; Kossinets and Watts 2006], or by the interaction of nodes connected by links [Gilbert and Karahalios 2009].

**Homophily (similarity)** This property isolates target users that are similar to local users. Various metrics may be used to calculate similarity, including structural properties such as degree and friend lists. Alternately, similarity in "taste" or content consumption patterns may also be used. One application of similarity is to identify a subset of links that are highly similar and therefore optimal for the given application needs. Where needed, similarity between arbitrary pairs of nodes can also be used to add or infer links in social networks. For example, Sripanidkulchai et al. [2003] constructs social networks for peer-to-peer systems based on their similarity of interest.

**Trustworthiness** This property identifies target users who are trusted by the local users. In some systems, trust is seen as a binary value, and the existence of a social link is seen as an indicator of trust. In other cases, trust may be captured as a numeric value; various methods such as the number of common friends [Granovetter 1985], the number of hops on the shortest path connecting two nodes [Golbeck 2006] have been used as measures of trust.

**Transitivity (clustering)** Transitivity considers social structure of three nodes and shows the friends-of-friends relationship. Recommending friends-of-friends to be friends is a common method for user selection in friend recommender systems, used by LinkedIn [LinkedIn 2008] and Facebook [Ratiu 2008], for example.

## 6.2 Group Delimitation

Group delimitation is another method to utilise social networks. Different from user selection, group delimitation selects target users according some general principles, without considering (or personalising for) specific local users. Thus, instead of focusing on individual nodes, social properties here are used to describe the common features of groups. It is usually used to detecting users with some common characteristics, such as spammers and Sybil users.

Group delimitation methods typically use a pattern learned from history. This involves checking each user and classifying users satisfying different patterns into different groups. The patterns are typically derived from topological properties that show the general (structural) features of nodes, for example, *transitivity*, *community* and *fast-mixing*. In addition, some *numerical* or *categorical* values extracted from local properties may also be used in group selection, such as *link semantics*, *centrality*, *average clustering coefficient* and *average closeness*.

Based on these social properties, node patterns can be *observed* or *inferred* from the activity history within system applications:

**Observed pattern** Most systems designers manually identify system-specific patterns and use them in their applications:

**Transitivity (Clustering)** For example, in anti-spam systems Boykin and Roychowdhury [2005], make the observation that spammers send emails to a large number of unrelated people, and can be expected to have an unreasonably low number of replies. Thus the clustering coefficient of spammers should be higher than legitimate users. This *observed* pattern is used to categorise nodes into spammers and legitimate users.

**Community** Community-based approaches delimit users belonging to a given community. For example, Khambatti et al. [2004] propose search methods in peer-to-peer systems where queries are flooding within the interest communities.

**Fast-mixing** Fast-mixing is an important property for Sybil detection (§9.2.2), which helps to decide whether an user is in the Sybil region or not without knowing all the members of that region.

**Link semantics** The idea is to delimit users of a given category. This method can be useful in some systems. For example, Yang et al. [2012b] propose category-specific social trust circles for recommendation systems, based on the observation that users will trust different users in different domains.

**Inferred pattern** A more sophisticated approach to find patterns is to use machine learning approaches. For example, Lam and Yeung [2007] use email senders' social properties to train a classifier which classifies emails into spam or ham.

## PART II: APPLICATIONS

In recent years, numerous systems have been proposed to make use of the properties of social networks. Having seen the properties in the first part of this article, we now survey three main application domains –recommender systems (§7), content-sharing systems (§8) and anti-abuse systems (§9). In each case, we survey representative examples that cover most of the different ways in which we see social properties being used, discuss how they fit within the framework proposed in §6, and point out areas for potential research. Content recommendations, content sharing and trust computations are fairly common requirements in a wide range of systems; thus our choice of application domains was influenced by the fact that some systems designers may be able to directly use some of the methods described herein without any modifications, in a "plug-and-play" manner. The three application domains were also chosen to illustrate how application needs may impact the use of social properties: The literature on recommender systems (§7) illustrates how to limit the scope of searches and interest matching using social context. Content sharing systems (§8) need to reconcile this social scope with the fact that large content items create heavy traffic and therefore content needs to be shared locally. Anti-abuse systems (§9) need to solve the interesting problem of limiting the scope of interactions to "honest" nodes in the system and thereby exclude potential abusers, without directly knowing who the honest and abusive nodes are.

## 7 RECOMMENDER SYSTEMS

Recommender systems collect information on the preferences of their users for a set of "items", and then provide them with predictions and recommendations of new "items". Initially, social networks were explored in recommender systems as yet another source of recommendation information, because users are known to like items liked by their friends. This has been shown quantitatively in several studies. For example, Lerman [2007] performed a large-scale empirical analysis on a social news aggregator, Digg, and showed that users tend to like stories submitted, read, or liked by friends. A similar study [Lee and Brusilovsky 2010] implemented on a collaborative tagging system, Citeulike, shows that users connected by social networks exhibit significantly higher similarity on their items, meta-data, and tags than non-connected users. Recent recommender systems are not just for items in the traditional sense. Some new applications, e.g., recommendations for new friends, locations and tags, have emerged.

In this section, we will discuss how social information is incorporated in both item and friend recommendations, followed by a discussion on the trend of research in this area.

## 7.1 Item Recommendations

Collaborative filtering is the most widely used method for recommender systems [Bobadilla et al. 2013; Tang et al. 2013]. It is based on the principle that users would like items that are liked by users with similar preferences. Recommender systems incorporate features that allow users to express preferences about different items, e.g., by giving ratings. Collaborative filtering recommends new items for a given *local user*, based on preferences of other users (sometimes termed as the *recommending source*) who have the most in common with the local user. Social collaborative filtering expands on this, utilising social information to improve the recommending source. Instead of merely finding the most similar users (who may be strangers) as the source of information, social-based recommendation systems use link, trust and tie strength provided by social graph, as discussed below.

*7.1.1 Links.* The idea of link-based methods is that users like items liked by their friends. For instance, Victor et al. [2009a] consider a user's immediate friends in the social graph as the recommending source, and recommends items liked by friends of a given user. SOFIA [Dell'Amico and Capra 2008] is an extension of traditional collaborative filtering: it recommends items based on items preferred by users who are similar to a given local user, but restricts the recommending source to the set of users who are declared as friends of the local user. This makes it more robust against sybil and other attacks which are possible in traditional collaborative filtering systems [Mobasher et al. 2007].

*7.1.2 Trust.* Many social recommender systems incorporate trust into their recommendations based on the intuition that users are likely to be influenced by friends they trust [O'Donovan and Smyth 2005]. In fact, Massa and Avesani [2007a] show that the recommender system can achieve better performance when replace the finding similar users step with finding users with high trust metrics. Based on such observation, [Ma et al. 2009, 2008] use probabilistic graphical models that fuses the user-item matrix with the users' social trust networks, and show that such models can generate better recommendations than the non-social collaborative filtering algorithms.

A common method of these systems is that users with higher trust values for a given local user would be given greater weight as recommendation sources. But different systems may employ different methods to measure such trust values between users. For example, TidalTrust [Golbeck 2006] uses a distance-based metric to estimate trust values among users: the closer two users are in the social graph, the higher the trust value between them. Massa and Avesani [2007a] use both local trust metrics (e.g., MoleTrust [Massa and Avesani 2007b]) and global trust metrics (e.g., PageRank [Page et al. 1999])

Recently, based on the fact that users' content consumption may be influenced by their social connections [Crandall et al. 2008], Chaney et al. [2015] model users' content consumption into two factors, users' own preference and the influence of their friends. So they develop a probabilistic model to incorporate social network information into a traditional factorisation method. The model infers each user's preferences and influences, and subsequently recommends items related both to what a user is likely to be interested in and what her friends have clicked.

*7.1.3 Tie strength.* These systems are based on the assumption that taste similarity is likely to be higher between users with "stronger" ties than than between those with "weaker" ties (see

§6.1 for common definitions of tie strength). Some of these kind of systems [Arazy et al. 2009; Carmagnola et al. 2009; He and Chu 2010] are similar to link-based methods, adding tie strength as a weighting factor when considering links. Model-based social recommendation systems [Tang et al. 2013] assume an underlying model to generate ratings. Many such systems, e.g. [Yang et al. 2012b], typically use tie strength as a component in their models to make them more robust.

## 7.2 Friend Recommendations

The act of introducing or suggesting new friends to one another is a basic activity in social life, but methods to automatically recommend new friends using algorithms had not been widely studied until the emergence of social network sites. Traditional collaborative filtering is not suitable to be used in this area directly, because it may need a user-friend rating matrix (although early works such as FilmTrust [Golbeck 2006] have attempted this approach). The process of recommending friends is similar to the process of finding a recommending source in item recommending; a number of social properties can be used for this. Friend recommendation methods can be organised into three categories, based on the information used: similarity, friend-of-friend and interaction.

*7.2.1 Similarity.* This category of systems is based on the theory of homophily which assumes that users who are similar to each other are more likely to become friends. Therefore, several social properties are utilised to compare the similarity of users. Then those who are highly similar to target users will be recommended as friends. For example, Hsu et al. [2006] consider both group features (e.g., degree, number of mutual friends) and interest-based features (e.g. mutual interests) when calculating similarity. Twittomender [Hannon et al. 2010] finds several strategies to represent a user its similarity calculation in the context of Twitter: the users tweets, tweets of her friends, and the friend list of the user. Akehurst et al. [2012] use implicit features inferred from explicit preferences by using machine learning methods, and show that ranking based on implicit or inferred similarity is significantly more accurate than explicit similarity based on declared user preferences.

*7.2.2 Friend-of-friend.* This kind of system uses the target user's friends of friends as a recommending source. This is a common method, used by both LinkedIn [LinkedIn 2008] and Facebook [Ratiu 2008], for example. The main difference between systems in this category is how they rank the recommending sources, which usually is based on overlaps between friend lists of the target users and recommending source, as measured by features such as the number of mutual friends [Silva et al. 2010], number of reciprocated friends [Armentano et al. 2012], etc.

Several studies have compared the similarity and friend-of-friend methods. Chen et al. [2009], using a personalized survey, find both kinds of algorithms effective in expanding users' friend lists, but have different functions: friend-of-friend recommendations are "better-received", and help users discover which of their contacts share their interests; algorithms using similarity of user-created content are better for discovering new friends. Similarly, Brzozowski and Romero [2011] find that in HP's enterprise social network, WaterCooler, sharing an audience with a user is a more compelling reason to follow that user, than straightforward similarity of interests to that user. This highlights the role of social contacts not only in helping to find interesting content but also in *shaping* the interests of users.

*7.2.3 Interaction.* Interaction is another feature considered in friend recommending. Zhang et al. [2011] explore how interaction (reply, comment) affects the establishment of friendship in Slashdot. They find that the higher the reply frequency, the higher the probability that a friendship would be established. CollabNet [Cai et al. 2012] uses the collaborative filtering method to consider friend recommendation. Users are considered to have a dual role as both "users" and "items". Then a matrix is constructed according to the interaction with active users, and a collaborative filtering approach is used to recommend new users.

| Systems | | Local | | | | | Topological | | | Method |
|---|---|:---:|:---:|:---:|:---:|:---:|:---:|:---:|:---:|:---:|
| | | Link | Degree | Tie strength | Homophily | Trustworthiness | Transitivity | Community | Fast-mixing | **Method** |
| *Item recomm.* | [Victor et al. 2009a] | √ | | | | | | | | U |
| | SOFIA [Dell'Amico and Capra 2008] | √ | | | √ | √ | | | | U |
| | TidalTrust [Golbeck 2006] | √ | | | | √ | | | | U |
| | [Massa and Avesani 2007a] | √ | | | | √ | | | | U |
| | SOREC[Ma et al. 2008] | √ | | | | √ | | | | U |
| | [Ma et al. 2009] | √ | | | | √ | | | | U |
| | [Chaney et al. 2015] | √ | | | | √ | | | | U |
| | [Arazy et al. 2009] | √ | | √ | √ | | | | | U |
| | [Carmagnola et al. 2009] | √ | | √ | | | | | | U |
| | [He and Chu 2010] | √ | | √ | | | | | | U |
| | [Li et al. 2015] | √ | | | √ | | | | | U+G |
| *Friend recomm.* | [Hsu et al. 2006] | √ | | | √ | | √ | | | U |
| | Twittomender [Hannon et al. 2010] | √ | | | √ | | | | | U |
| | [Akehurst et al. 2012] | | | | √ | | | | | U |
| | [Silva et al. 2010] | | | | √ | | √ | | | U |
| | [Armentano et al. 2012] | | | | √ | | √ | | | U |
| | CollabNet [Cai et al. 2012] | | | | √ | | | | | U |
| | [Brzozowski and Romero 2011] | | | | √ | | | | | U |
| | [Yang et al. 2012b] | √ | | | √ | | | | | U+G |
| | TopRec [Zhang et al. 2013] | √ | | | √ | | | | | U+G |
| | [Backstrom and Leskovec 2011] | | | √ | | √ | | | | U |

Table 2. **Comparison of recommender systems**, where "U" indicates using user selection method, and "G" indicates group delimitation method

### 7.3 Discussion

From the above survey for recommender systems, it is clear that social networks are utilised widely for user selection. For item recommendation, recommending sources are detected, and for friend recommending, potential friends are explored. We organise these systems into Table 2 according to the social properties that they have used. From the table, we find that almost all friendship-based user selection properties are utilised in this area, but a few group delimitation approaches are also explored. Further research into using the group delimitation methods could be a promising potential direction for this area.

Similarly, several different kinds of social networks (e.g., offline, interaction-based, online and heterogeneous) have been utilised for recommendations. But the usage of heterogeneous social networks still needs to be studied further, as discussed below.

## 7.4 Potential for future work

Although there has been an explosion of interest on the social-based recommendation recently, some aspects have still not received enough attention:

*7.4.1 Delimiting groups.* As our brief survey showed, most studies consider the problem of user selection in recommender systems; studies that use group delimitation is limited. Group delimitation can improve the accuracy of user selection and/or improve the quality of the recommending source. A few recent studies have explored how to delimit the source of recommendation information using item categories. For example, Yang et al. [2012b] try to infer category-specific social trust circles from available rating data combined with social network data and gets better accuracy. TopRec [Zhang et al. 2013] uses a semi-supervised probabilistic topic model to mine communities and corresponding topics from the social network. The detected topical communities are used to construct interpretable domains for domain-specific collaborative filtering. Recently, Li et al. [2015] also allow overlapping communities of social networks and incorporate them into personalised item recommendations. This is a significant step forward as real world communities are often overlapping.

*7.4.2 Heterogeneity in social networks.* Although most existing recommender systems consider social links homogeneously [Tang et al. 2013], as we discussed in §2.4, social links are heterogeneous and represent various types of relationships in reality. For instance, a user who is on both Facebook and Twitter may have different friends on each. Even within, say, Twitter, a user may have professional as well as personal contacts; some followers might even be strangers, who are interested in the user's tweets. The category-specific recommendation [Yang et al. 2012b] that we just discussed above can be seen as specialising for user heterogeneity, but other information can also be utilised. For example, aggregating social graphs from different data sources: SONAR [Guy et al. 2008] tries to aggregate social relationship from different public data sources for recommendation, while Social-Union [Symeonidis et al. 2011] is a method that combines rating matrices from several different social rating networks. Future research could potentially benefit from combining inferred social links with social links declared explicitly by users on traditional social networks.

*7.4.3 Link prediction.* Another trend in the area is to model the recommendation as link prediction to make the problem more general. For example, Backstrom and Leskovec [2011] proposed a supervised random walk-based link recommendation method. It uses node and edge features to learn edge weights such that the random walk on the weighted network is more likely to visit "positive" nodes, nodes that have a higher probability of creating a new edge to the starting node of the random walk.

*7.4.4 Incorporating negative links.* While the vast majority of recommender systems utilise *unsigned* social networks, i.e., only considering positive social links, some recent studies also take negative social links (refer §4.2) into consideration. For example, Tang et al. [2016a] propose a recommender model that consider both positive and negative social links, based on the observation that when a user has both friend (positive) and foe (negative) circles, the user's preference is more likely to be closer to that of her friend circle than that of foe circle. So they add a penalty to the system and pull users' preference closer to friend circle during the recommendation process. Tang et al. [2016b] summarises three strategies to consider negative social links: (1) consider negatively linked users as "unwanted" users and avoid any recommendations from them [Victor et al. 2009b]; (2) consider negative links as dissimilarity measurements [Victor et al. 2013]; (3) In reality, social links in signed networks are sparse, so one idea proposed by Nalluri [2014] recently is to propagate positive and negative values in networks before actually considering them.

*7.4.5   Using specialised social networks.* Apart from generic online social networks, several specialised social networks have emerged for different purposes such as location sharing, music, collaborative tagging etc (§3.2). These can serve as high quality sources of recommendation information. For example, tag recommendation [Feng and Wang 2012; Rae et al. 2010; Shepitsen et al. 2008] is an active area for this research. The problem is complex because tags are an unstructured form of meta data where the vocabulary and reasoning behind each user's choice of tags varies. At the same time, recommending tags for music, images and tweets may need different methods. Some studies consider item-user-metadata networks as three heterogeneous, linked structures and help alleviate the cold start problem caused by data sparsity [Feng and Wang 2012]. Applications may be tailored for particular kinds of information even within a single website, for instance, in Twitter, different methods have been developed for recommending tweets [Yan et al. 2012], mentions [Wang et al. 2013] and hashtags [Kywe et al. 2012; Zangerle et al. 2011]. Location-related recommendation [Bao et al. 2013; Carretero et al. 2012; Karamshuk et al. 2013; Scellato et al. 2011b] is another active area, which is based on the checkin information provided by location-based social networks, e.g., Gowalla and Foursquare. Location information has also been used to recommend new items, such as venues [Noulas et al. 2012], and Ads [Saez-Trumper et al. 2012].

## 8   CONTENT-SHARING SYSTEMS

In contrast to recommendations which can be treated as content metadata, content-sharing systems are infrastructure to support delivery of content. Social network-based systems in this area can be classified into peer-to-peer file sharing networks, where peers work co-operatively to distribute content amongst themsleves, or content distribution networks, which consist of a geographically distributed set of servers which maintain replicas of content items close to the users who issue requests. Both have in common that they involve a community of users who generate requests for shared data. Small world patterns and some other properties have been verified in their data-sharing graphs [Iamnitchi et al. 2004, 2011; Wang et al. 2006], which has inspired researchers to construct social networks in those systems to improve their performance. System considerations differ slightly depending on whether the intended application is elastic file downloads (§8.1), where content discovery and optimal use of any available bandwidth are important, or real-time video streaming (§8.2), where reliable delivery of video chunks in time for playout during a video stream is key.

### 8.1   Peer-assisted elastic file downloads

Peer-to-peer (P2P) systems are platforms for sharing and exchanging resources among thousands or even millions of users. In P2P systems, data is transferred directly between computers belonging to two end users of the network, or two *peers*. Content search and discovery in such systems is difficult, because information about content availability is distributed; often there is no central catalogue of all the information. Several different approaches have been tried to solve this problem [Tigelaar et al. 2012]; one approach is to use information embedded in social networks. The basic assumption is that the social neighbours of a peer are more likely have the content she is querying, so it can be efficient to send queries to social friends before flooding it to all peers. But P2P users usually do not explicitly declare their social networks, so social network-based P2P systems have to infer social networks according to user interests or connect to online social networks (Figure 3), as discussed below.

*8.1.1   Inferred social networks.* Most social network-based systems in P2P networks use inferred social networks to avoid changing the basic structure of P2P systems. They are based on the theory of homophily, which assumes that users that are similar to each other have a higher probability of being friends. So inferred social networks link users similar to each other and the similarity is usually calculated according to previous communication history. For example, Sripanidkulchai et al. [2003] use query history to calculate the similarity and assume that if a peer has a particular piece of content that one is interested in, it is likely that she will have other items that one is interested in as well. So the sender and responder of each successful query will create shortcuts to one another, and preferentially send queries to each other in following search periods. These shortcuts provide a loose structure on top of traditional unstructured overlays, resolving queries quickly and dramatically reducing the total load in the system. Although they have not defined social networks in their paper, the shortcuts overlay is similar to social networks used in other P2P systems. Tempich et al. [2004] create a social network according to query traces, and attempt to propose a routing strategy that uses historical successful query records to improve content querying.

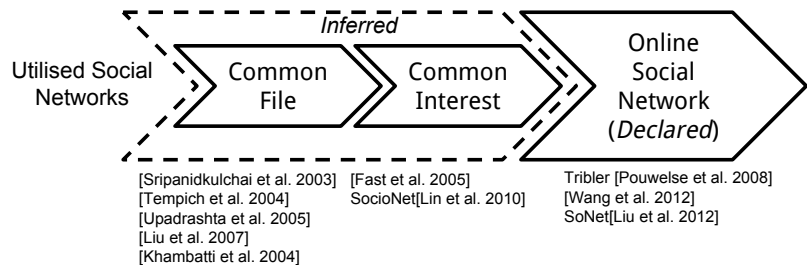

Fig. 3. **The construction of social networks in peer-to-peer systems**: Early systems relied on inferred links between users based on common files; subsequent systems mined relationships based on common interests, and more recently explicitly declared friendships on online social networks have been used.

Upadrashta et al. [2005] present a system in which nodes maintain friend lists learnt from historical query results. But a node may have different friend lists for different semantic areas (domains), so queries can be routed semantically to them. Liu et al. [2007] have similar idea. Each peer in their system maintains a number of tables that associates each topic with the most relevant peers. Hence each peer can forward the query to the most relevant peers of the requesting object by looking up these tables. Khambatti et al. [2004] propose a slightly different system which limits search scope to communities formed by peers with common interests.

Some systems also try to infer interest of peers, so that users who have no overlap in their query histories, but are interested in similar files, are connected. For example, Fast et al. [2005] propose to construct social links according to the *type* of files (obtained through clustering files in the network) that users have shared. SocioNet [Lin et al. 2010] is a social-based overlay that clusters peers based on users' preference relationships as a small-world network and thus assists peers in locating content at peers with similar interest through short path lengths. The evaluation on AudioScrobbler[10] shows that SocioNet achieves a higher success ratio compared with other methods, while reducing message overhead significantly.

8.1.2 *Declared social networks.* Tribler [Pouwelse et al. 2008] is another example of a social network-based P2P system. Instead of constructing the social network via common interests, it allows users to explicitly make friends with each other, in a way similar to other online social networks. Through maintaining such a social network, it enables a significant improvement in download performance. Tribler introduces a collaborative download protocol, 2Fast, which allows users to invoke the help of their friends to form trusted groups who altruistically co-operate to speed up each others' downloads. Peers in a social group join a swarm and contribute bandwidth even though they are not interested in the content shared in the swarm, and get better download performance through the help of their friends. 2Fast introduces the notion of social group incentives to encourage users to donate bandwidth.

---

[10] AudioScrobbler, a database tracks users' music listening habits, has merged with last.fm in 2005, according to Wikipedia (http://en.wikipedia.org/wiki/Last.fm, last accessed on Jun 15, 2017).

Instead of constructing new social networks, some recent studies explore how to directly use external online social networks to improve P2P performance. For example, Wang et al. [2012] show that an increasing number of torrents are being shared on Twitter (supported in part by popular clients such as uTorrent, which allows users to talk about torrents using Twitter). They observe that two peers who participate in a normal BitTorrent [Cohen 2003] swarms meet again with a probability lesser than 5%, whereas in Twitter-triggered swarms, peers meet again with more than 35% probability. Based on this, they propose a simple modification to BitTorrent leechers' unchoke algorithm, prioritising unchokes from social friends who have the highest uploading rate, and find that this improves the download completion time significantly.

In addition to unstructured P2P overlays, social properties have also been explored in structured P2P overlays. SPROUT [Marti et al. 2005] is a social network-based method upon a structured P2P network that relies on a Distributed Hash Table [Stoica et al. 2001]. In addition to the links of the structured DHT overlay, SPROUT leverages existing social network services to establish additional, highly trusted, links cheaply. It prefers forwarding messages through social friends to avoid routing messages through possibly malicious nodes. Yet another application is clustering users based on social networks, for more efficient and trustworthy content discovery. For example, SoNet [Liu et al. 2012] incorporates a hierarchical distributed hash table overlay to cluster nodes based both on common interests and on proximity, and connects nodes with social links. In this way, it enables friendship-aware intra-subcluster querying and locality- and interest-aware intra-cluster searching.

## 8.2 Video-streaming systems

Online video sharing systems, such as YouTube and vimeo, have changed user habits of video consumption significantly. Since those services usually embed a social network, social relationships are explored to improve those systems as well. Furthermore, it has been observed that social networks account for a large fraction of the audience for many videos [Broxton et al. 2010]. Thus, delivery infrastructure can optimise for the viral popularity of videos in social networks. Most of recent studies focus on two categories of systems: peer-assisted video-streaming and content delivery networks.

*8.2.1 Peer-assisted video-straming systems.* The first category of studies focuses on "peer-assisted" systems, which borrow ideas from P2P systems to enable data transfers among peers participating in video-stream. NetTube [Cheng and Liu 2009] is a such system, which explores the clustering in social networks for short video sharing. It supplies an efficient method to locate a peer's next video using the social network and a social-based pre-fetching strategy which allows peers to download prefixes of related videos to reduce video startup latency. Wang et al. [2011] design a pre-fetch strategy based on users' preferences learnt from their historical videos access patterns. In particular, users' preferences are predicted by the source user of the videos, the social distance to the source user and the popularities of the videos. Xu et al. [2015] propose a hierarchical community structure for mobile multimedia streaming services. The structure is based on social relationships extracted from user contact from online social network and real society, considering user mobility and interest similarity. Their simulation results demonstrate that systems with such structure can achieve high content sharing efficiency and QoS levels.

| Systems | | Local | | | | | Topological | | | Method |
|---|---|---|---|---|---|---|---|---|---|---|
| | | Link | Degree | Tie strength | Homophily | Trustworthiness | Transitivity | Community | Fast-mixing | |
| Elastic file downloads | [Sripanidkulchai et al. 2003] | √ | | | √ | | | | | U |
| | [Tempich et al. 2004] | √ | | | √ | | | | | U |
| | [Upadrashta et al. 2005] | √ | | | √ | | | | | U+G |
| | [Liu et al. 2007] | √ | | | | | | | | U+G |
| | [Khambatti et al. 2004] | √ | | | √ | | | √ | | U+G |
| | [Fast et al. 2005] | √ | | | √ | | | | | U |
| | SocioNet [Lin et al. 2010] | √ | | | √ | | | | | U |
| | Tribler [Pouwelse et al. 2008] | | | | | √ | | √ | | U+G |
| | [Wang et al. 2012] | √ | | | | | | | | U |
| | SPROUT [Marti et al. 2005] | | | | | √ | | | | U |
| | SoNet [Liu et al. 2012] | √ | | | √ | | | | | U |
| Video streams | NetTube [Cheng and Liu 2009] | | | | √ | | | √ | | U+G |
| | [Wang et al. 2011] | √ | | | √ | | | | | U |
| | [Xu et al. 2015] | √ | | | √ | | | √ | | U+G |
| | [Sastry et al. 2009] | √ | | | | | | | | U |
| | [Scellato et al. 2011a] | √ | | | | | | | | U |
| | [Wang et al. 2012] | √ | | | | | | | | U |
| | [Traverso et al. 2012] | √ | | | | | | | | U |

Table 3. **Comparison of content-sharing systems**, where "U" indicates using user selection method, and "G" indicates group delimitation method

*8.2.2 Content delivery networks.* The second category of systems apply social information to enhance traditional Content Delivery Networks (CDNs), which work by replicating content from an origin server to different locations and redirecting user requests to locations close to them. Localising flows this way eliminates potential bottlenecks at the origin server and also makes it easier to guarantee better quality of service for users. This global replication strategy is efficient for the most popular content, but it may not practical for items in the tail of the popularity distribution, since they are accessed too rarely. So the problem is how to distribute content among local servers. Sastry et al. [2009] and Scellato et al. [2011a] explore the use of social cascades and geographic information to provide hints for the replicas placement, while Wang et al. [2012] also take the propagation of content into account. Tailgate [Traverso et al. 2012] proposes a method based on the access patterns exacted from social relationships and time-zone differences, to selectively and efficiently distribute tail content in a lazy just-in-time manner based on predicted times of access.

## 8.3 Discussion

In Table 3, we summarise social network-based content sharing systems according to the social properties they have utilised. In the table, we show that those systems are mainly dependant on personalised user selection methods, especially link-structure and similarity-based methods. Group delimitation is also used by them to limit query flooding.

Further, since content-sharing systems are based on communication networks and sometimes predate the popularity of online social networks, many social-based methods construct (infer) their own social networks rather than using declared pre-existing social links. However some recent attempts show that external online social networks can also be used to improve content delivery and sharing.

### 8.4 Potential for future work

This section considered P2P and content replication systems. Lots of work has been done in these areas, and also in using social information in each area individually. However, we believe there is scope to add social recommendations to improve P2P systems. And similarly, as OSNs scale, and incorporate more content within themselves [Beaver et al. 2010], they face problems in content delivery, especially because of their centralised nature. So, designing replication for OSNs, or designing decentralised social networks in a secure manner become interesting problems, which are starting to be addressed. In this section, we will discuss future work of content-sharing systems from those aspects.

*8.4.1 Integration with other application areas.* Content discovery and recommendation services (discussed in §7) are also explored by some content-sharing systems recently. For example, Tribler [Pouwelse et al. 2008], that we have introduced in previous section, provides such services through linking users with similar tastes and periodically exchanging preference lists among those users. More specifically, using an epidemic protocol called Buddycast, each peer in Tribler maintains a number of taste buddies with their preference lists and a number of random peers in their Megacache. Then periodically, peers either send a Buddycast message exchange request to one of its taste buddies (exploitation) or to a random peer (exploration), or replies to such message received from another peer. A Buddycast reply message contains a number of taste buddies along with their top-$N$ preference lists, a number of random peers and the top-$M$ (usually, $M > N$) preferences of the sending peer. After exchanging Buddycast messages, both peers merge the information in the message received into their own Megacache, so that a recommendation can be performed according to it. Similarly, Vera-del Campo et al. [2012] propose an epidemic routing method to recommend documents according to the userâĂŹs interests. Draidi et al. [2011] attempt to recommend high quality content related to query topics through exploiting by useful friends (of friends) of the users in social networks. A peer becomes useful to another peer if their interests overlap.

*8.4.2 Support for online social network systems.* It is natural that properties of social networks could be used to support online social network systems. The primary issue here is that social data typically involves many-many communication and is both produced and consumed within the social network itself (For example, consider Facebook Wall posts or email communications, where the data, i.e., wall post or email, is produced within a social context, and is consumed by intended recipients of the communication, who are also users of that social network). This creates a high degree of inter-dependency, which makes it difficult to distribute data storage across multiple locations.

[Wittie et al. 2010] shows that Wall traffic on Facebook is mostly localised between friends in the same geographic region. Thus, by introducing regional TCP proxies, both the load on Facebook's central servers as well as the latency of user access can be reduced. [Karagiannis et al. 2010] and [Pujol et al. 2010a] consider data from a company's email network and Twitter respectively and show that users can be partitioned efficiently using versions of the METIS[11] graph partitioning tool. [Silberstein et al. 2010] show how to deliver the latest data and improve performance at the same time, by selectively materialising the views of low-rate producers and querying the latest data from high-rate producers.

---

[11]http://glaros.dtc.umn.edu/gkhome/views/metis, last accessed on Jun 15, 2017.

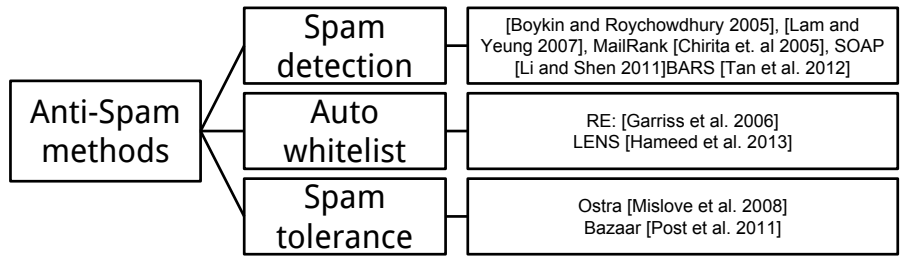

Fig. 4. The classification of spam systems

Some recent systems attempt to improve the performance of online social networks using content delivery networks. Similar to previous systems, those systems also focus on scaling online social networks. For example, SPAR [Pujol et al. 2010b] optimises the social graph partition problem to achieve data locality and minimized replication number. S-CLONE [Tran et al. 2012] improves the replication process by taking into account social relationships.

*8.4.3   Decentralised online social networks.* To meet the privacy needs of online social networks, some recent studies propose to deploy them on decentralised structures, which are based on or inspired by P2P systems. For example, Persona [Baden et al. 2009] uses decentralised, persistent storage so that user data remains available in the system and users may choose with whom they store their information. Similarly, Safebook [Cutillo et al. 2009] propose two design principles, "decentralisation and exploiting real-life trust". [Buchegger et al. 2009; Nilizadeh et al. 2012] also explore decentralised structure for online social networks. In general, there are four design issues that each decentralised online social network architecture would need to address [Nilizadeh et al. 2012]: the storage of social structure and creation of node-node links (e.g., whether to use a structured distributed hash table or use the social graph and link socially connected nodes); the location of content shared by users (in owner's machine, in their friends' machines, or across the system, accessed using a distributed hash table method); content dissemination method (pushed by producers, or pulled by consumers as required); and access control (cryptographically enforced, or using online authentication). An important motivation for decentralised approaches is that centralised approaches typically require trusting a server to with highly sensitive social data. However, additional research is required to attain the same degree of scale and efficiency possible in current centralised social networks.

## 9   ANTI-ABUSE SYSTEMS

In this section, social-based methods in two anti-abuse systems: anti-spam and anti-sybil systems, will be reviewed and how they utilise social properties are explored. We find they tend to use generalised user selection methods.

### 9.1   Systems to combat spam

Spam defence schemes are usually employed in email systems, where social networks can be built according to the communication history and fields in email headers like "From", "To", "Cc" and "Bcc". Generally, spam defences can be divided into three categories (Figure 4): spam detection, auto whitelist and spam tolerance, as we discuss below.

*9.1.1   Spam detection.* Traditional spam filtering mechanisms try to divide emails into valid emails ("ham") or spam according to some patterns of the email content or the senders of the email. But social-based methods construct social networks according to communication history first and then use social properties to describe the patterns. For example, Boykin and Roychowdhury [2005] note that spammers use temporary email addresses and send emails to a number of unrelated people. Thus, triangle structures (§4.1) are largely absent in their subnetworks in the email social graph. So the absence of clustering, or a low clustering co-efficient, is used to label spam. Lam and Yeung [2007] use a machine-learning approach to semi-automatically classify spam based on a number of social network characteristics such as user degree and clustering co-efficient. Another kind of approach seeks to use trustworthiness and reputation to decrease spam, but these methods also try to turn those properties into some patterns. For example, in MailRank [Chirita et al. 2005], the reputation score of each email address is calculated according to the trustworthiness of other users. All addresses are ranked according to reputation score, and low ranked ones will be considered as spammers.

Instead of constructing social networks based on email communication patterns, some recent studies utilise social relations in online social networks. For example, SOAP [Li and Shen 2011] is a online social network-based spam filter that integrates social closeness, user (dis)interest and adaptive trust management into a Bayesian filter. Cao et al. [2015] is a friend spammer detect system in online social networks. It is based on the observation of social rejections, i.e., even well-maintained fake accounts inevitably have their requests rejected or they are reported by legitimate users. Rejecto uses this insight to partition the social graph into two regions such that the aggregate acceptance rate of friend requests from one region to the other is minimised. This leads to reliable detection of the region that comprises friend spammers.

*9.1.2 Auto whitelist.* Whitelist is another spam defence strategy. It allows users to specify some trusted addresses, so that email from those addresses will pass the spam filter without checking. Recent studies try to build the whitelist automatically using social networks. RE: [Garriss et al. 2006] is an example. It is based on the intuition that friends-of-friends are unlikely to be spammers and automatically whitelists them. However, if a user Alice wishes to communicate with another user Bob, Bob may not wish to expose the list of his friends to Alice directly. To discover whether there is an intersection of friend lists, RE: introduces a novel zero knowledge protocol and verifies user's friends and friends of friends using secure attestations. Because a majority of legitimate emails originate close to the receiver's position in the email social network, they demonstrate, using real email traces, that up to 88% of real emails which would have been mis-classified as spam, can be identified. LENS [Hameed et al. 2013] is a similar system but based on online social networks. It leverages the recipient's social network to allow correspondence within the social network to directly pass to the mailbox of the recipient. To enable new senders to send emails, some legitimate and authentic users, called Gatekeepers, are selected from outside the recipientâĂŹs social circle and within predefined social distances. These gatekeepers provide tokens attesting that a user is not a spammer, enabling whitelisting beyond friend-of-friend circles.

*9.1.3 Spam tolerance.* The third strategy of defending spam is spam tolerance. Those schemes usually are based on *credit networks* [DeFigueiredo and Barr 2005] which enables transitive trust based on credit tokens. In credit networks, to receive services from others, nodes need to pay credits for it. In spam tolerance methods, the actions that an adversaries could take by creating Sybil accounts (see §9.2.1) is limited.

A good example of a credit-based spam tolerance system is Ostra [Mislove et al. 2008], loosely based on Hawala, the ancient Indian system of debt value transfers (still prevalent in India and a few countries in Asia and Africa). Ostra prevents spam by requiring a chain of credit transfers in the social network to send emails. Sending emails decreases the available credit balance. This balance is eventually restored if the recipient accepts an email. This system allows legitimate senders to operate freely, but sending bulk emails (or falsely marking as spam) quickly pushes the sender (or a lying recipient) against the credit limits, and prevents further damage. Note that Ostra is not overtly based on any social network property, but relies on nodes cutting links to untrustworthy spammers in order to preserve their own credit lines. Bazaar [Post et al. 2011] is a successor, which introduces a max-flow computation and secures reputations on online marketplaces from Sybil and whitewashing attacks.

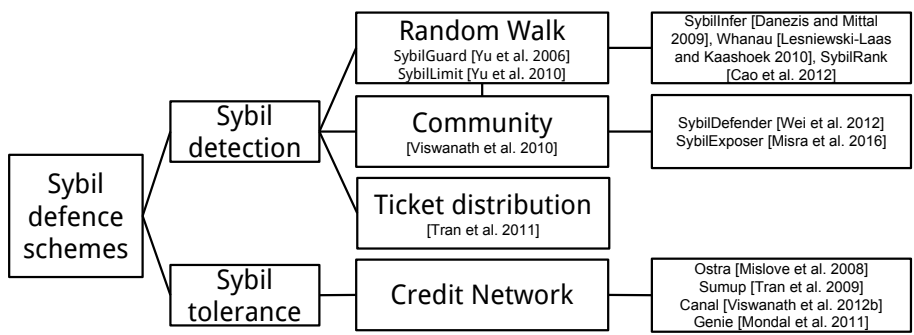

Fig. 5. The organisation of sybil systems

## 9.2 Systems to combat Sybil attacks

Sybil defence schemes can be divided in to two categories [Viswanath et al. 2012a, 2010] (Figure 5): *Sybil detection* and *Sybil tolerance*. The former category of approaches attempt to distinguish Sybil nodes from honest ones, while the latter category merely tries to bound the benefits that adversaries can gain after Sybil attacks.

*9.2.1 Sybil detection.* The target of Sybil detection is to label users as either honest or Sybil with high accuracy, which is a typical generalised user detection problem. Those systems are based on the following central assumption [12]: While an adversary can create any number of Sybil identities and relationships between them, it is much harder to *induce* an *attack edge* between a Sybil node and an honest one. Nodes that on one side of the set of attack edges is the set of honest nodes, or the *honest region*; on the other side, the set of Sybil nodes, or the *Sybil region*. Most Sybil detection methods try to use the these topological features to partition the network into honest and Sybil regions.

In general, there are two major approaches for Sybil detection. The first [Yu et al. 2010, 2006] relies on performing multiple random walks on the social network. The random walks are unlikely to cross attack edges because they are scarce. However, random walks from two nodes on the same side of the attack edge can easily intersect because they do not have to cross the attack edge. By using a large enough number of random walks from each node, the probability of intersection can be made high enough for two nodes in the same region. Thus, the central protocol involves a verifier node and a suspect node. The verifier accepts a suspect node if their random walks intersect. Performing such random walks is typically assumed to be easy because social networks are considered to be fast-mixing (see §5.2).

SybilInfer [Danezis and Mittal 2009] is a variant of this random walk-based approach. It creates a trace based on random walks, and then, accepts or rejects new sets of nodes using the trace and a Bayesian model which decides if the nodes belong to the Sybil or honest region. Another variant is

---

[12]However, some empirical studies, such as [Yang et al. 2012a, 2011], find this assumption may be inappropriate.

Whanau [Lesniewski-Laas and Kaashoek 2010], which develops a Sybil-proof Distributed Hash Table based on random walks. SybilRank [Cao et al. 2012] performs some short random walks from an honest node and then ranks nodes according to the degree-normalised probability. Low-ranked nodes are considered as suspects.

Interestingly, Viswanath et al. [2010] analyse a number of social network-based Sybil defences and show that they all work by detecting local communities. Following this discovery, some community-based Sybil defence approaches have been proposed. For example, SybilDefender [Wei et al. 2012] presents a two-step approach which identifies Sybil node first and then detects Sybil community around it. SybilExposer [Misra et al. 2016] expands the idea of [Yu et al. 2006] to the community-level and is built upon the intuition that there are more inter-community links between honest communities than from honest communities to Sybil communities. But Alvisi et al. [2013] warn that the choice of community detection protocol requires extreme caution, as it dramatically affects the Sybil detection result. The results of Viswanath et al. [2010] also suggest that social graphs with well-defined communities are more vulnerable and that adversaries which can create targeted links can be more effective.

The second approach to sybil detection is based on the realisation that when attack edges are scarce, the maximum flow between the honest and Sybil regions is small [N.Tran et al. 2011]. Instead of actually computing the flow, the approach uses a ticket distribution mechanism: starting from each verifier node, tickets are distributed in a breadth-first manner (which is similar with SumUp [Tran et al. 2009], that we will discuss in §9.2.2), and only those nodes that have enough tickets will be accepted by the protocol. If a verifier node is close to the Sybil region, then Sybil nodes could end up with some tickets. By carefully choosing multiple verifier nodes randomly across the network, the protocol removes this sensitivity to the positions of individual verifiers.

*9.2.2 Sybil tolerance.* Instead of separating Sybil nodes from honest nodes, Sybil tolerance systems seek to bound the leverage that adversaries could gain from doing that. Ostra [Mislove et al. 2008], which we have discussed in §9.1 could be seen as a Sybil tolerance systems.

SumUp [Tran et al. 2009] is a Sybil tolerance system for online content rating. In this system, credits are distributed along links in the network and voters must find a path with available credits to the selected vote collectors in order to provide a rating. Most honest users can participate, because there are plenty of trusted paths from the honest user to a vote collector (due to fast mixing). But for the Sybil nodes, since links between them and honest nodes are limited, most cannot find a path to the vote. Canal [Viswanath et al. 2012b] is an extension of SumUp, that uses landmark routing techniques decrease the computational cost of credit payments. Genie [Mondal et al. 2011] is also a credit network-based system that can defend against Sybil crawlers.

## 9.3 Discussion

Table 4 shows that anti-abuse systems mainly rely on user-selection methods to identify and isolate users abusing a given system. Breaking this down according to the taxonomy of Figure 4 and Figure 5, we see that: (1) group delimitation is highly utilised in all types of anti-abuse systems. (2) Some spam *detection* methods tend to use methods based on (inferred or observed) pattern matching. By contrast, sybil *detection* schemes tend to use random walk-based methods to delineate groups of sybil identities to avoid; this relies on properties of link structure and trustworthiness of links—random walks would easily cross over into Sybil regions if large numbers of links existed between honest, trustworthy nodes and dishonest, sybil nodes. (3) When it comes to abuse *tolerance* schemes, both spam and Sybil tolerance systems are similar, since both try to limit the benefits that abuse accounts can obtain.

| Systems | Local | | | | | Topological | | | Method |
|---|---|---|---|---|---|---|---|---|---|
| | Link | Degree | Tie strength | Homophily | Trustworthiness | Transitivity | Community | Fast-mixing | |
| **Spam defence** | | | | | | | | | |
| [Boykin and Roychowdhury 2005] | | | | | √ | | | | G |
| [Lam and Yeung 2007] | | √ | | | | √ | | | G |
| MailRank [Chirita et al. 2005] | | | | | √ | | | | G |
| SOAP [Li and Shen 2011] | | | | | √ | √ | | | G |
| RE: [Garriss et al. 2006] | √ | | | | | | | | G |
| LENS [Hameed et al. 2013] | √ | | | | | | | | G |
| Ostra [Mislove et al. 2008] | √ | | | √ | | | | | G |
| Bazaar [Post et al. 2011] | √ | | | √ | | | | | G |
| **Sybil defence/tolerance** | | | | | | | | | |
| SybilGuard [Yu et al. 2006] | | | | | | | | √ | G |
| SybilLimit [Yu et al. 2010] | | | | | | | | √ | G |
| SybilInfer [Danezis and Mittal 2009] | | | | | | | | √ | G |
| Whanau [Lesniewski-Laas and Kaashoek 2010] | | | | | | | | √ | G |
| SybilRank [Cao et al. 2012] | | √ | | | | | | √ | G |
| SybilDefender [Wei et al. 2012] | | | | | | | √ | √ | G |
| SybilExposer [Misra et al. 2016] | | | | | | √ | | √ | G |
| [N.Tran et al. 2011] | | | | | | | | √ | G |
| SumUp [Tran et al. 2009] | | | | | | | | √ | G |
| Canal [Viswanath et al. 2012b] | | | | | | | | √ | G |
| Genie [Mondal et al. 2011] | | | | | | | | √ | G |

Table 4. Comparison of anti-abuse systems

## 9.4 Potential for future work

*9.4.1 Which social network to use?* As discussed previously, a main concern for this class of social systems has been whether social links are as trustworthy as they are claimed to be, and whether properties attributed to social networks (e.g., Fast Mixing) hold generally in all social networks. We discuss below how using different social networks might confer new benefits to social systems. As background, we note that traditionally, anti-spam systems have tended to use email communication-based social networks whereas anti-sybil systems implicitly or explicitly assume online social networks. However, recently, anti-spam systems also attempt to connect with and use online social networks. Similarly, anti-sybil systems may also benefit from constructing communication-based networks and thus getting a heterogeneous social structure.

Systems such as SOAP [Li and Shen 2011] and LENS [Hameed et al. 2013] are early examples that attempt to utilise social networks constructed in online social websites. Another trend is to extend anti-spam systems to detect spam on social websites, or *social spam* [Heymann et al. 2007]. For example, Lee et al. [2010] try to automatically harvest spammers' profiles and tweets in Twitter through a honeypot-based approach. Then they identify some specific features (e.g., content, friend information, posting pattern) of spammers and use them to train a classification algorithm. Similarly, BARS [Tan et al. 2012] proposes a runtime spam detection scheme based on behaviour patterns of spammers in social websites. Benevenuto et al. [2010] find that the fraction of followers per followees in Twitter could distinguish spammers from regular users. These systems illustrate the benefit that can be obtained from cross-fertilising the social networks used between different applications.

*9.4.2 Crowdsourcing methods.* Crowdsourcing abuse detection is another new direction in security systems, where the abuse detection work is outsourced to an undefined group. For example, Wang et al. [2013] conduct a user study and designs a system that scales to millions of profiles. As a proof-of-concept, it shows that provided the right seed set of expert users, the system is effective. Another similar method is allowing users to report abuse users or provide blacklist functions, similar to sites such as Malware Patrol[13] or PhishTank[14]. In this case, social links could be used to provide additional context, or for making the system more effective. For instance, negative social relationships may be incorporated and properties such as structural balance can be used to identify personally relevant URLs that should either be allowed or disallowed (e.g., A system may disallow adult URLs to underage users by expanding their parental blacklists based on similar filters constructed for their friends by their friends' parents).

## 10   RELATED WORK

With the explosion of research interest in social networks, a large body of literature has appeared in this area, including several surveys. These can be divided into *surveys about social networks*, and *surveys about other topics that include social network-related aspects*.

Surveys about social networks can in turn be classified into those that take a *generalist* view, and those that have a *specialist* focus on more specific aspects. Generalist surveys typically examine a wide variety of social networks and tend to cover topics such as the definition and history of social networks [boyd and Ellison 2008], or their evolution [Leberknight et al. 2012]. Berger et al. [2014] provides a kind of citation analysis, studying aspects such as what the major research areas have been, and which journals have been most receptive to social networks papers.

In contrast with this generalist approach, many surveys focus deeply on certain *kinds* of social networks. For example, Kayastha et al. [2011] and Karam and Mohamed [2012] examine mobile social networks, systems which provide a variety of data delivery services involving the social relationship among mobile users. Bao et al. [2013] review some related studies on location-based social networks, while Gupta et al. [2010] and Mezghani et al. [2012] focus on social tagging systems. Tang et al. [2016b] review systems incorporating signed social networks. and provide a research survey of information systems on social networks.

Another class of specialist approach is to focus on some specific aspect of social networks. For instance, Sherchan et al. [2013] present a review of literatures on trust in social networks. Yet another specialist angle is to focus on some topic of interest such as the use of machine learning and computational networks used in social computing [King et al. 2009].

Our approach falls between the specialist and generalist perspectives: we are generally interested in all aspects of social networks and their properties, but with a specific focus on how they can be applied in or used by some real-world application such as security, recommendation or content sharing. This leads to the two parts of the survey, with the first introducing a large range of properties, and the second discussing specific application areas.

Individually, each of the systems design areas we consider has had several surveys, including some which consider social network-informed designs: for instance [Bao et al. 2013; Bobadilla et al. 2013; Tang et al. 2013] on recommender systems, [Androutsellis-Theotokis and Spinellis 2004; Pathan and Buyya 2007; Tigelaar et al. 2012] for content-sharing systems and [Caruana and Li

---

[13]http://www.malware.com.br, last accessed on Jun 15, 2017.
[14]http://www.phishtank.com, last accessed on Jun 15, 2017.

| Systems | Local | | | | | Topological | | | Method |
|---|---|---|---|---|---|---|---|---|---|
| | Link | Degree | Tie strength | Homophily | Trustworthiness | Transitivity | Community | Fast-mixing | |
| Label [Hui and Crowcroft 2007] | | | | | | | √ | | U+G |
| Bubble Rap [Hui et al. 2008] | | √ | | | | | √ | | U+G |
| LocalCom [Li and Wu 2009] | | | | | | | √ | | U+G |
| SocialCast [Costa et al. 2008] | | | | | | | √ | | U+G |
| SANE [Mei et al. 2011] | | | | √ | | | | | U |
| [Wu and Wang 2012] | | | | √ | | | | | U |
| [Bigwood and Henderson 2008] | √ | | | √ | | | | | U |
| [Bulut and Szymanski 2010] | √ | | | | | | | | U |
| [Zhang and Zhao 2009] | √ | | | √ | | | | | U |
| SimBet [Daly and Haahr 2007] | √ | | | √ | | | | | U |
| PeopleRank[Mtibaa et al. 2010] | √ | √ | | | | | | | U |
| [Abdelkader et al. 2010] | √ | | √ | | | | | | U |
| [Gao and Cao 2011] | | √ | | √ | | | | | U |
| [Fabbri and Verdone 2011] | | √ | | | | | | | U |
| Give2Get [Mei and Stefa 2010] | | | | | √ | | | | U |
| TFT [Shevade et al. 2008] | | | | | √ | | | | U |
| SSAR [Li et al. 2010] | | | | | √ | | | | U |

(Left vertical label: Delay tolerant networks)

Table 5. Methods of delay tolerant networks

2008; Heymann et al. 2007; Hoffman et al. 2009; Yu 2011] for anti-abuse systems. Although some of these surveys do consider social network-based methods, they do so within the context of the particular kind of system (for example, recommender systems). Our larger perspective allows us to draw up a taxonomy of usage *across* application areas, showing that most applications of social network properties are for user selection or group delimitation (§6).

We chose the above four classes of systems as representative showcases for illustrating how social properties can be applied to different tasks. There are other systems applications that readers may be interested in. For example, social networks have been utilised for message routing in delay tolerant networks, a class of mobile communication systems for transmitting messages in a store-carry-forward fashion, in communication scenarios where infrastructures are destroyed or non-existent [Karamshuk et al. 2011; Wei et al. 2013; Zhu et al. 2013]. However, we note that our framework applies in this class of systems as well, as summarised in Table 5. We hope that the generality of our framework will assist systems designers in applying social properties to other classes of applications, where social networks have not been considered before.

## 11 CONCLUSION

In this work, we attempt to give a broad overview of the utilisation of social networks in system applications. We proposed a framework (Figure 1) that divides social network-based methods into *user selection* and *group delimitation*. First, we surveyed social properties that may be utilised in system applications and discussed how they could be employed for user selection and group delimitation methods. Then, we selected three typical classes of system applications: recommender systems, content-sharing systems and anti-abuse systems. We categorised each class of systems into a unifying framework according to the social properties utilised. For each class of systems, we propose potential directions for future work based on gaps identified in the application of social properties. To make it easier for incorporating social information in a variety of settings, we also provided a brief overview of how social links are inferred or collected in different application contexts and online social network platforms, and also discussed potential pitfalls to be aware of in using certain claimed properties such as trustworthiness of social links.

We chose the above three classes of systems as representative showcases for illustrating how social properties can be applied to different tasks. We conclude by noting that our framework has been designed to be general enough to apply to other classes of systems. For example, an important class of systems applications which uses social properties is delay-tolerant networks, where data may be routed over social contacts in a store-carry-forward fashion, constructing paths over time even when there is no contemporaneous end-to-end path between the source of the message and its destination. Several surveys exist for such systems, for example, [Karamshuk et al. 2011; Wei et al. 2013; Zhu et al. 2013]. Table 5 illustrates how our framework applies to this class of systems as well.

## ACKNOWLEDGMENTS

The work was partially supported by the UK Economic and Social Research Council (ESRC) under Grant No.: ES/M00354X/1 (http://www.space4sharingstudy.org/) and the UK Engineering and Physical Sciences Research Council (EPSRC) under Grant No.: EP/K024914/1.